\documentclass[12pt, preprint]{aastex}
\newcommand\ea{et al.\ }

\def\chandra{{\it Chandra}}
\newcommand\psc{\ifmmode{\rm\,cm^{-2}}\else{${\rm\,cm^{-2}}$}\fi}
\newcommand\kms{\ifmmode{\rm\,km\,s^{-1}}\else{${\rm\,km\,s^{-1}}$}\fi}
\shorttitle{Cygnus Loop SE Knot}
\shortauthors{Levenson and Graham}

\begin{document}
\title{Environmental Impact on the Southeast Limb of the Cygnus Loop}

\author{N. A. Levenson\altaffilmark{1} and James R. Graham\altaffilmark{2}}
\altaffiltext{1}{Department of Physics and Astronomy, University of Kentucky,
Lexington, KY 40506; levenson@pa.uky.edu}
\altaffiltext{2}{Department of Astronomy, University of California, Berkeley, CA 94720; jrg@astron.berkeley.edu}

\begin{abstract}
We analyze  observations from the {\it Chandra X-ray Observatory} 
of the southeast
knot of the Cygnus Loop supernova remnant.  
In this region, the blast wave propagates through an 
inhomogeneous environment.  
Extrinsic differences and subsequent multiple
projections along the line of sight
rather than intrinsic shock variations, such as fluid
instabilities,
account for the apparent complexity of the images.
Interactions between the supernova blast wave and density enhancements
of a large interstellar cloud 
can produce the  morphological and spectral characteristics.
Most of the  X-ray flux arises in such interactions, not in
the diffuse interior of the supernova remnant.
Additional observations at optical and radio wavelengths support this
account of the existing interstellar medium and its role in shaping
the Cygnus Loop, and they demonstrate that the southeast knot is
not a small cloud that the blast wave has engulfed.
These data are consistent with rapid equilibration of electron and
ion temperatures  behind the shock front, and the current blast
wave velocity $v_{bw} \approx 330 \kms$. 
Most of this area does not show strong evidence for non-equilibrium 
ionization conditions, which may be a consequence of  the high 
densities of the bright emission regions.
\end{abstract}
\section{Introduction}
Supernova remnants play a  fundamental
role in processing  matter and energy in galaxies.
Supernovae create the heaviest elements, and their
remnants mix the enriched ejecta into the
surrounding interstellar medium (ISM).
Supernova remnants (SNRs) are the primary source
of the hot component of the ISM, and the supernova rate,
with the remnants' subsequent evolution,  
determines whether this component predominates in
any particular galaxy.
The extant environment reciprocally  affects the 
evolution of a SNR.
If large, dense clouds populate
the surroundings with large covering fraction,
the blast wave's expansion will stall, 
limiting the amount of thermal energy that is injected into the ISM.

The Cygnus Loop is a middle-aged ($8,000$ years old) supernova
remnant, now interacting with large-scale inhomogeneities in
the surrounding ISM.  These variations 
are responsible for its observable characteristics, including
a limb-brightened X-ray shell and strong correlations between
X-ray and optical surface brightness.
Propagating through the low-density environment its progenitor
evacuated, the undisturbed blast wave has a velocity 
$v_s \approx 350 \kms$ and produces low surface brightness X-ray 
emission with temperature $kT \approx 0.15$ keV
($T \approx 1.8\times 10^6$ K). 
Upon encountering a large, dense cloud, however, the
blast wave is decelerated, to $v_s < 200 \kms$.  The post-shock material
rapidly cools through optical line emission 
to $T \approx 10^4 $ K. 
An important consequence of the cloud interaction is the development
of reflected shocks.  These shocks propagate back through the hot, compressed
interior of the SNR, enhancing X-ray emission \citep{Hes86}.

The northeastern  and western limbs of the Cygnus Loop offer
two clear examples of these interactions between 
shock fronts and large clouds \citep*{Hes94,Lev02}.
Although less prominent than these bright regions,
the southeast knot is physically similar.
\citet*{Fes92} first drew attention to this feature,
and subsequent work by  \citet{Gra95} demonstrated that it 
represents  the early stage of the encounter between the
blast wave and a large interstellar cloud.
The southeast knot is also located near a portion of
the primary blast wave traced in H$\alpha$ at optical
wavelengths.  
These are ``nonradiative'' shocks, which excite
Balmer line emission through electron collisions in unshocked gas
that is predominantly neutral \citep{Che78}.  Thus, they identify
sites where the gas is being shocked for the first time.
The shocked  gas has not yet had time to cool radiatively,
so the optical 
spectra lack the emission lines that
are typical of fully radiative post-shock regions, such as 
[\ion{O}{3}] and [\ion{S}{2}], in addition to the Balmer lines.

The Cygnus Loop is nearby and bright,
allowing high spatial resolution and 
high signal-to-noise observations
to investigate variations and shock evolution on
physically-relevant scales.
At the 440 pc distance of this SNR
\citep{Bla99}, $1\arcsec$ corresponds to a physical size of $6.6\times10^{15}$ cm.
Here we analyze new observations of the Cygnus Loop's
southeast knot obtained
with the  {\it Chandra X-ray Observatory (Chandra)}.
In addition to superb spatial resolution,
\chandra{} also provides simultaneous
spectroscopic information
for direct and unambiguous measurement of
the distinct physical character of different areas.
Combining these X-ray  data with previous optical  and radio
measurements, we distinguish 
temporal evolution of the post-shock medium
from external variations in the environment. 

The three-dimensional geometry of the Cygnus Loop
complicates the interpretation of observations. 
Multiple projections of the current blast wave location or
cloud interactions  can
emerge along a single line of sight.  As a result, 
areas that appear projected onto the interior of the SNR
may actually represent newly-shocked material that is
located in the foreground (or background), not
older shocked gas that is genuinely inside the hot SNR interior.
In the southeast, we illustrate that many of the observed variations
are a consequence of these projection effects.

\section{Observations and Data Reduction}
\begin{figure}
\includegraphics[width=\textwidth]{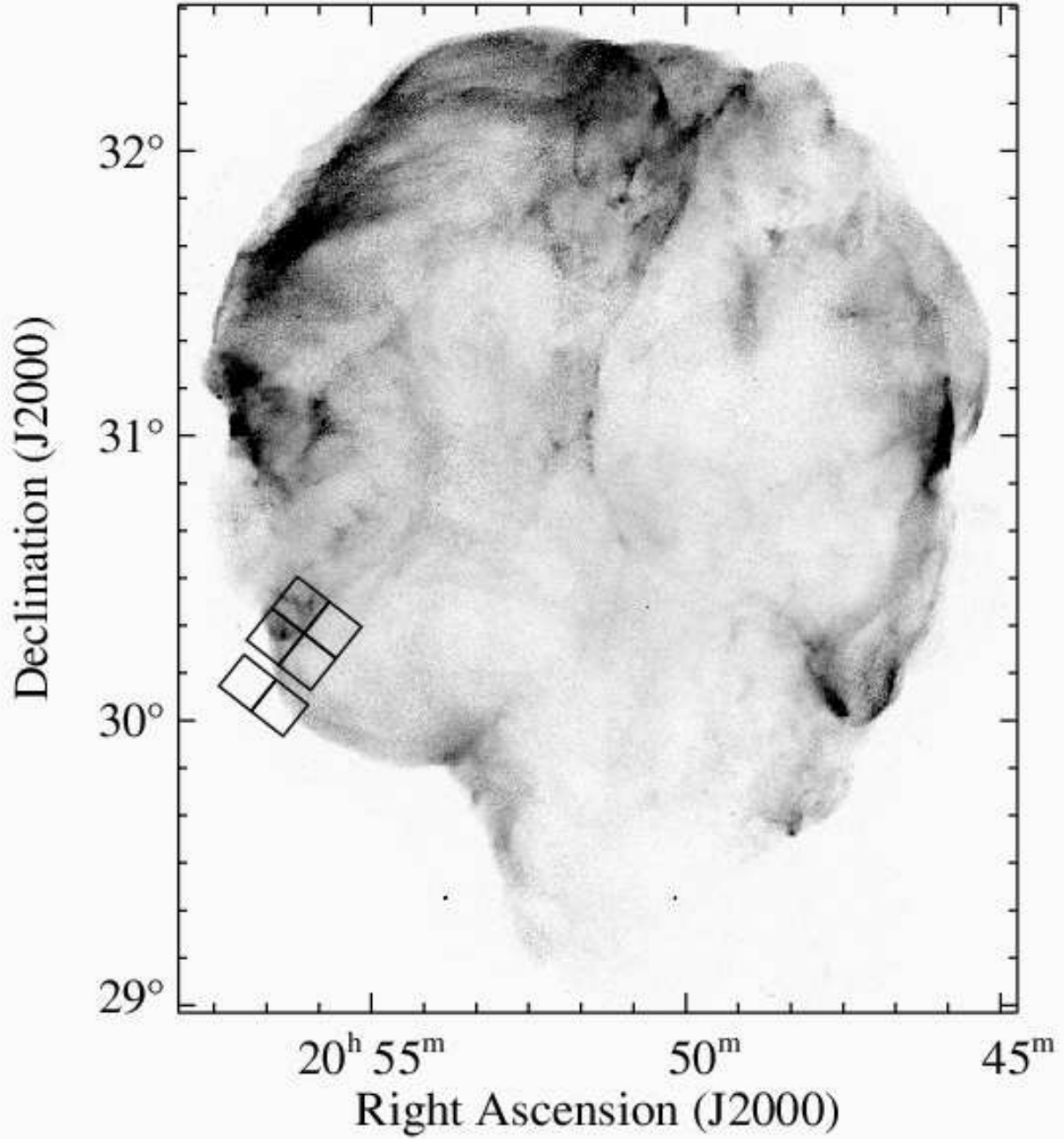} 
\caption{\label{fig:hri}
ACIS field of view overlaid on mosaic of {\it ROSAT} HRI
soft X-ray image \citep[completed with later observations]{Lev97}.
}
\end{figure}

We observed the southeast knot of the Cygnus Loop
with the  \chandra{} Advanced CCD Imaging Spectrometer 
(ACIS) on 2000 September 1,   
obtaining spatial resolution around $1\arcsec$
over energy $E = 0.3$--$8$ keV, and 
simultaneous spectral resolution $E/\Delta E \approx 12$.
The total field of view of the six CCDs is
approximately $25\arcmin \times 17 \arcmin$ and covers the
brightest X-ray  region of the southeast knot, some of the diffuse interior,
and some slight X-ray enhancements close to the blast wave.
Figure \ref{fig:hri} illustrates these observations in
the context of the Cygnus Loop as a whole.  The bright southeast knot
is at the apex of a slight indentation from the near-circular boundary of the SNR. 
Five of the CCDs (I0, I1, I2, I3, and S2) 
are front-illuminated (FI), having readout electronics 
that face the incident photons.  The sole back-illuminated (BI) CCD, S3, 
is located entirely outside
the blast wave boundary.  Because the sensitivity of the BI and FI
CCDs is different, the S3 data are not useful for measuring
background emission, so we will not discuss them.

We reprocessed all data from original Level 1 event files using Chandra Interactive
Analysis of Observations (CIAO) software, version 3.1, 
removing the 0.5-pixel spatial randomization that is included in standard
processing.
(See the \chandra{} Science Center\footnote{http://cxc.harvard.edu/} 
for details about \chandra{}
data and standard processing procedures.)
We applied current calibrations (Calibration Database version 2.28), 
including corrections for
charge-transfer inefficiency and time-dependent gain variations, to produce
the Level 2 event file.
We included only good events that 
do not lie on node boundaries, where discrimination of cosmic rays
is difficult.
We examined the lightcurves of background regions and found no
significant flares, so we did not reject any additional data
from the standard good-time intervals, for a net exposure
of 47 ks.

\begin{figure}
\includegraphics[width=0.9\textwidth]{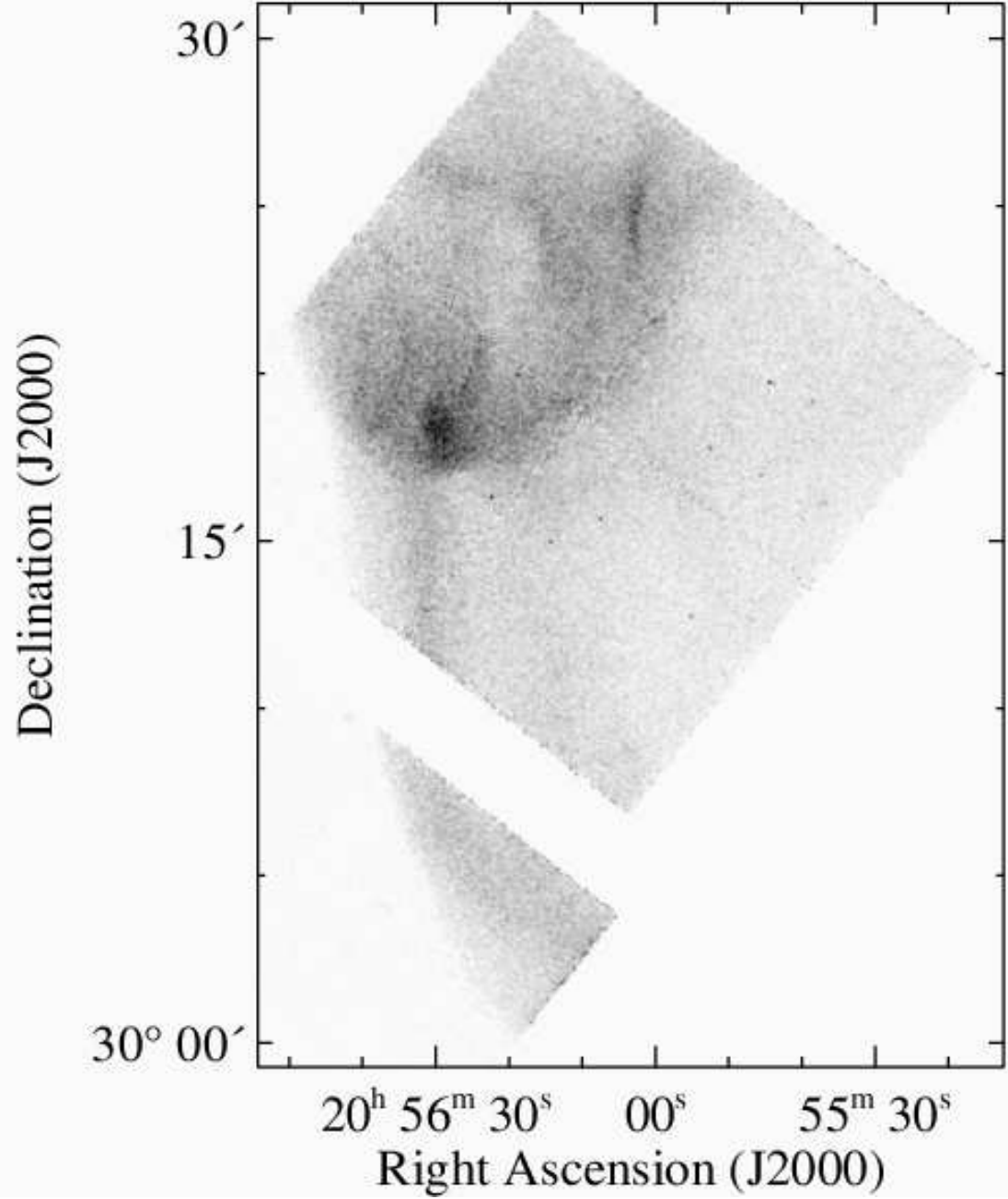} 
\caption{\label{fig:totimg}
Total (0.3--8 keV) exposure-corrected image, scaled
linearly from 0 (white) to 
$7\times 10^{-7}{\rm \, photons\, cm^{-2}\, arcsec^{-2}}$ (black).
The large gap that separates the four northern 
ACIS detectors from the southernmost one is unobserved.
}
\end{figure}

We created exposure maps to correct soft (0.3--0.6 keV), medium (0.6--0.9 keV),
hard (0.9--2 keV), and very hard (2--8 keV)
images separately.  The total exposure-corrected 
image (Figure \ref{fig:totimg}) includes all these bands,
with the raw pixels binned by a factor of two (to 1 arcsec$^2$)
and smoothed by a Gaussian of FWHM=5$\arcsec$.
Spacecraft dither during the exposure is sufficient to cover the small gaps
among the four northern CCDs, but a large gap separating the southern
CCD remains unobserved.

\begin{figure}
\includegraphics[width=0.9\textwidth]{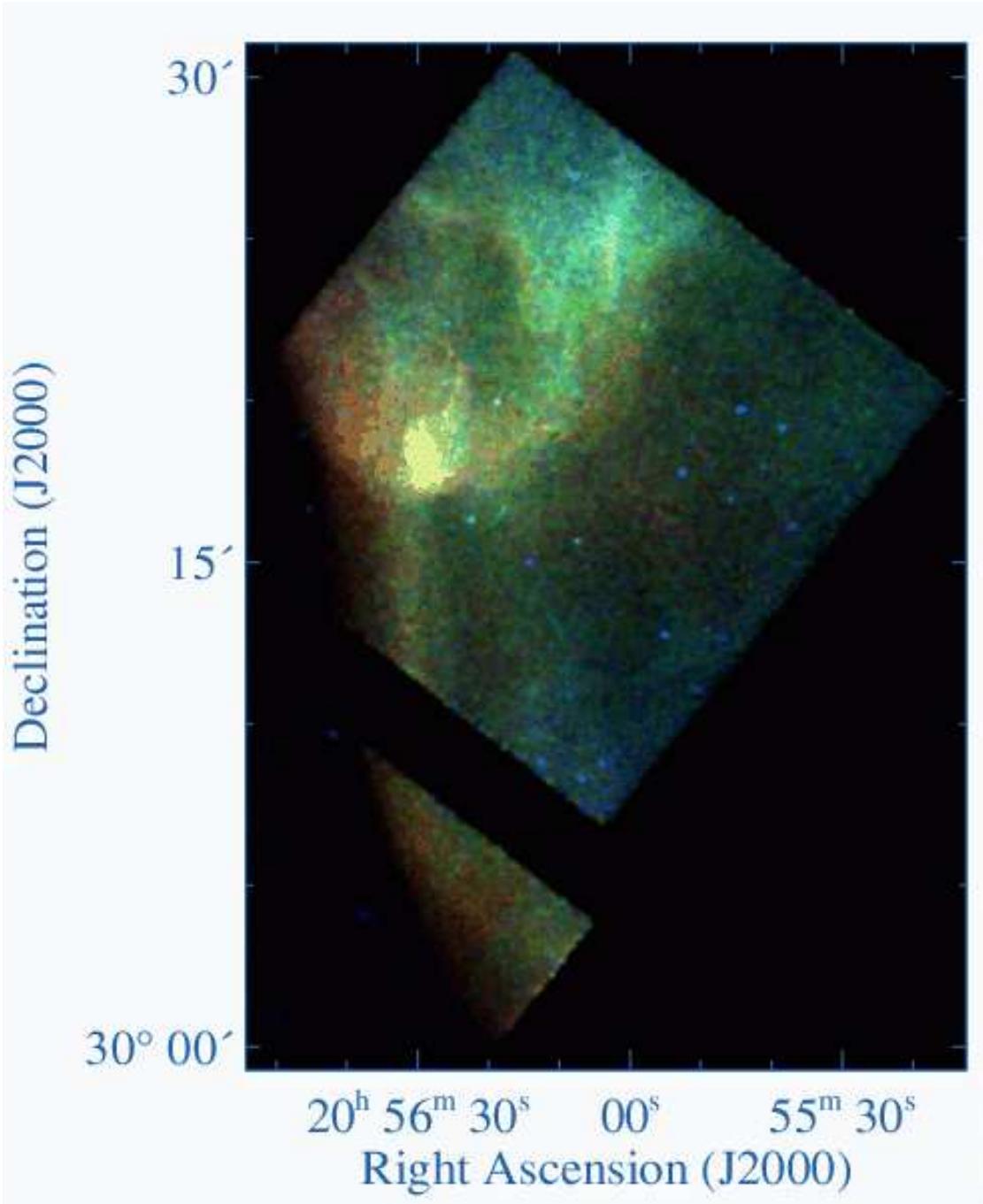}  
\caption{\label{fig:clrimg}
False-color image, constructed from 
soft (red; 0.3--0.6 keV), medium (green; 0.6--0.9 keV), and hard
(blue; 0.9--2 keV) exposure-corrected images.
The individual images have been smoothed by a Gaussian of
FWHM=5$\arcsec$ and scaled linearly, from 0 to
4, 1, and 0.3 $\times 10^{-7}{\rm \, photons\, cm^{-2}\, arcsec^{-2}}$,
respectively.
}
\end{figure}

Spectral variations are apparent in a false-color image
constructed from the three softer bands (Figure \ref{fig:clrimg}).  
Nearly all  the emission
above 2 keV is due to background point sources unlikely to be related to 
the SNR, and many of these hard sources are evident in the
blue band of the color image.  
The emission near the blast wave is generally soft,
and the low surface brightness interior is harder.
Stronger hard emission also arises near the site of interaction
with the dense southeast cloud.  

\section{Spectral Modeling}
\begin{figure}
\includegraphics[width=\textwidth]{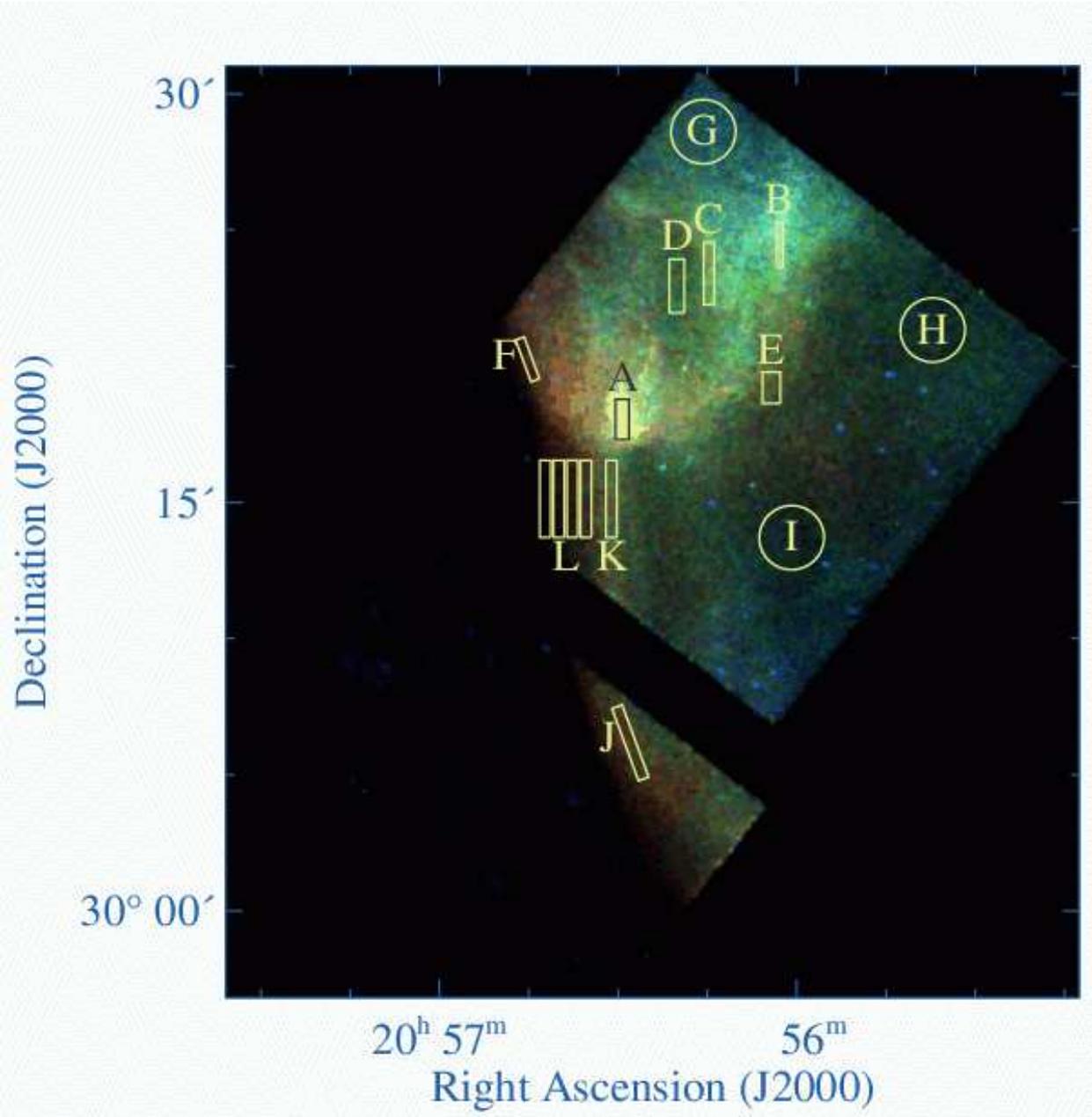} 
\caption{\label{fig:regs}
Spectroscopic apertures are identified on the 
false-color \chandra{} image.
}
\end{figure}

We select regions for spectral analysis from the color image,
identifying extended areas of distinct spectral character.
Figure \ref{fig:regs} shows these locations. 
The regions are not of uniform size or shape, 
in order to isolate unique physical characteristics 
within each aperture. 
The spatially-limited spectral extractions require specific
processing and calibration measurements.
As we noted above, we applied corrections for the  charge-transfer inefficiency and 
gain variations that have developed over time in
creating the Level 2 events file.
Another significant change 
over the course of the \chandra{} mission is the reduced
soft X-ray sensitivity, likely the
result of build-up of material on the detector.
In constructing ancillary response files, we used the CIAO contamination
correction  to account for this variation in quantum efficiency.

Because the Cygnus Loop covers nearly the entire field of view, we
measure the background emission in comparable observations of the
blank sky. The blank sky background spectra 
have been filtered and processed the same way as
the data\footnote{http://cxc.harvard.edu/contrib/maxim/acisbg/},
and scaled by the relative exposure time.
A portion of the I3 and S2 detectors covers the exterior
of the SNR.  We compared local background measurements 
for source spectra within these detectors and found no significant
difference from the blank-sky backgrounds.  For consistency, we use
the blank-sky background measurements in all cases. 

We fit the spectra in XSPEC \citep{Arn96}, 
grouping the spectra  into bins of a minimum of 20 counts,
so the errors are normally distributed and 
$\chi^2$ statistics are appropriate. 
We use data from 0.5 keV (below which the calibration becomes uncertain)
to 2.0 keV, the maximum energy of any significant emission. 
In general, we expect to find thermal emission behind the
blast wave of this SNR.
We use the collisional ionization equilibrium models of 
Borkowski et al.
\citep{Ham83,Bor94,Lie95}, including dielectronic recombination
rates of \citet{Maz98}, the VEQUIL model in XSPEC.

We consider abundance variations, 
which we quote relative to solar values of \citet{And89}. 
At the typical temperature of this area of the Cygnus
Loop, oxygen and iron produce the strongest emission lines.
While groups of elements may be expected to vary the same way
if they are produced through the same modes of stellar 
evolution---with other alpha elements similar to oxygen and other
iron group elements similar to iron---other members of these
nucleosynthetic families
are in fact not at all significant in the observed 
spectra.  Therefore, we do not  include their variation
in the spectral modeling.

Oxygen is the only element whose abundance differs
from the solar value, and it is consistently
underabundant relative to
the \citet{And89} value of $(O/H)_\sun = 8.51\times10^4$.
Our findings are more typical of the
local ISM, however.  Within 500 pc of the Sun, 
\citet*{Mey98}, for example, find gas-phase
$(O/H) = 3.43\times10^4$,
correcting for updated transition probabilities 
\citep[cited in \citealt{And03}]{Wel99}.

\begin{figure}
\includegraphics[angle=270,width=\textwidth]{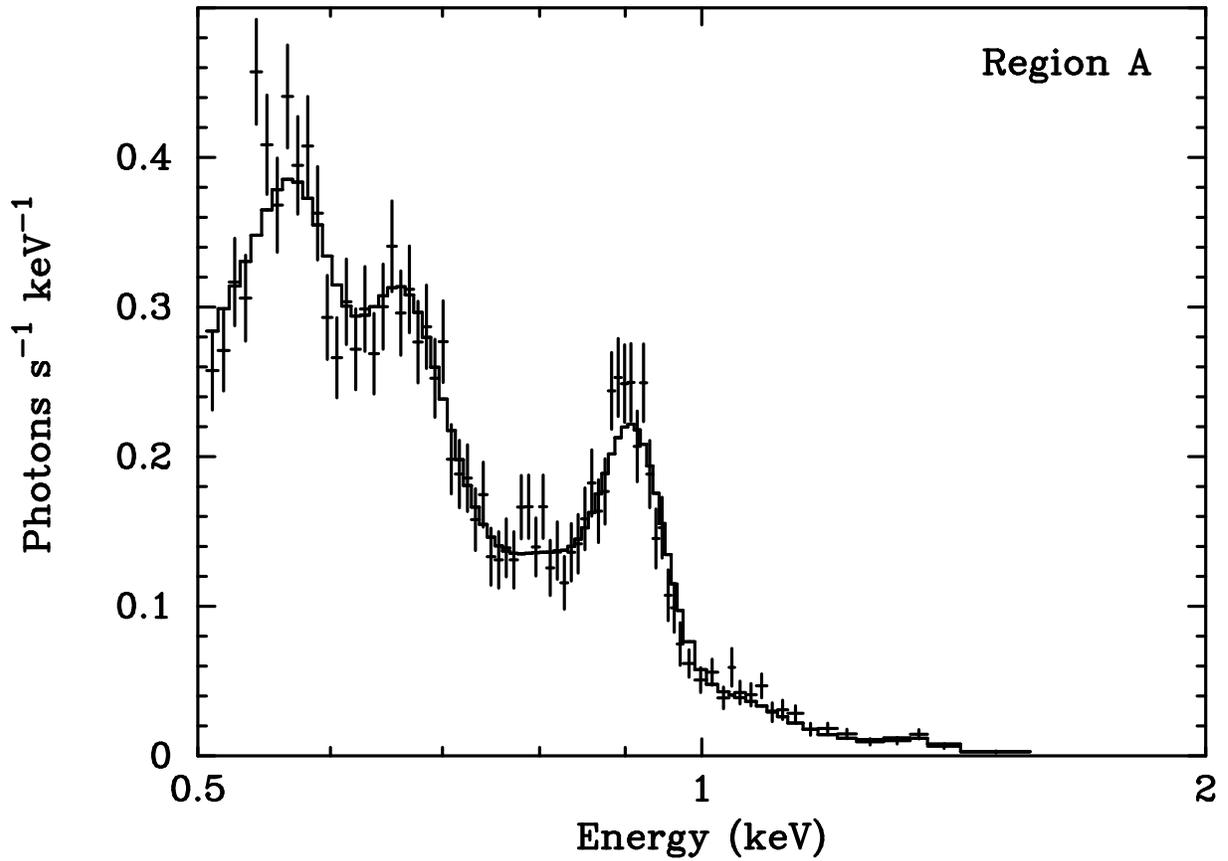} 
\caption{\label{fig:speca}
ACIS spectrum of region A (crosses) with the 
equilibrium model (histogram).
The thermal plasma is at temperature $0.17$ keV, absorbed by
$N_H = 1.0\times 10^{21}\psc$. 
}
\end{figure}

\begin{figure}
\includegraphics[angle=270,width=\textwidth]{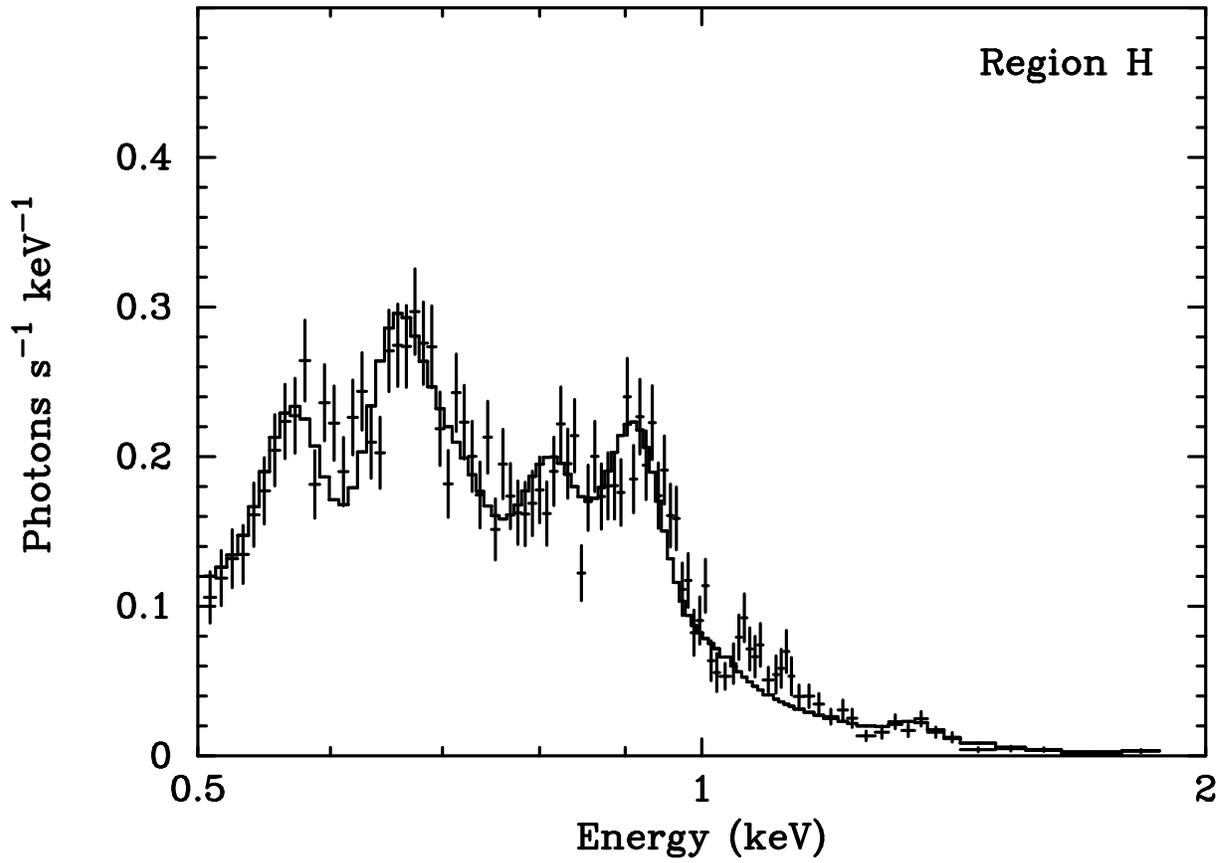} 
\caption{\label{fig:spech}
ACIS spectrum of region H (crosses) with the best-fitting
model (histogram), on the same scale as Figure \ref{fig:speca}.
The model includes two thermal equilibrium plasmas, at temperatures
0.16 and 0.57 keV, absorbed by $N_H = 2.1\times 10^{21}\psc$. 
}
\end{figure}

Some spectra demand more complex models to 
attain adequate fits, requiring additional temperature
components or  non-equilibrium conditions.
In the more complex models that are presented, 
the inclusion of additional model components and their
free parameters are significant at a minimum of the 95\% confidence
limit, based on an $F$-test.  All quoted errors are 90\% confidence
for one parameter of interest.
Figures \ref{fig:speca} and \ref{fig:spech} show two example spectra 
and their corresponding equilibrium model fits on the same scale.

\subsection{Equilibrium Results}

All single-temperature equilibrium model fits are listed in Table \ref{tab:spec}, 
with the values of
$\chi^2$ and the number of degrees of freedom (dof) in the last column.  
This model yields the best fits in regions B, D, E, and K.
In these regions, the single-temperature 
equilibrium plasmas have temperatures
$kT = 0.17$--$0.19$ keV
and sub-solar oxygen abundance.
These regions are generally associated with
the southeast knot and located near the shocked cloud's optical 
emission.  We discuss these spectral results in the context 
of this inhomogeneous environment below
(\S \ref{sec:interp}), after considering more complex models.

Regions G, H, and I  
require two thermal components, listed in
Table \ref{tab:spec2}.
The same foreground column density is applied to both components. 
These apertures 
are located in the  diffuse interior of the SNR, 
and they
are not directly associated with interactions
between the blast wave and the southeast cloud or the cavity boundary.
These areas are noticeably bluer in the energy-coded color image.
Each of these regions is fit
with a soft component similar to the single-temperature
regions, as well as a hotter ($kT_2 \approx 0.6$--$0.7$ keV)
component.
Qualitatively, we recognize that these apertures cover material that was heated
by the passage of the
original undisturbed blast wave, before it was strongly
decelerated in an encounter with higher-density material. 
At each location, shocked SNR material extends along the line of sight,
representing a range of original shock velocities.
In the two-temperature model, however, neither
component  quantitatively reveals
consistent adiabatic evolution of the blast wave.
We attempted to fit these spectra with a model that explicitly 
accounts for this evolution (the VSEDOV model), but
the resulting fits are worse.

\section{Non-Equilibrium Conditions}
\subsection{Non-Equilibrium Ionization}
In addition to the equilibrium descriptions above, we also examined 
non-equilibrium conditions that may be relevant in the Cygnus Loop.
One important  consideration is the ionization state relative
to the electron temperature.  Initially, the post-shock gas is
underionized compared with its equilibrium value.  In the spectrum, 
the temperature the lines indicate, based on ionization state,
then differs from the continuum temperature, which the electrons determine.
The  ionization timescale, $\tau = n_et$, parameterizes the degree
of equilibration in terms of $n_e$, the electron density, and 
$t$, the time elapsed since the
shock passage \citep{Gor74}.
The timescale for ionization equilibrium
varies for each element and is a function
of temperature.  For example, at $T = 2.0\times10^6$ K ($kT = 0.17$ keV),
the ionization equilibration $e$-folding timescale for oxygen 
$t_{ieq} \approx 3\times10^{11}n_e^{-1}$ s \citep{Lie99,Shu82}.

\begin{figure}
\includegraphics[angle=270,width=\textwidth]{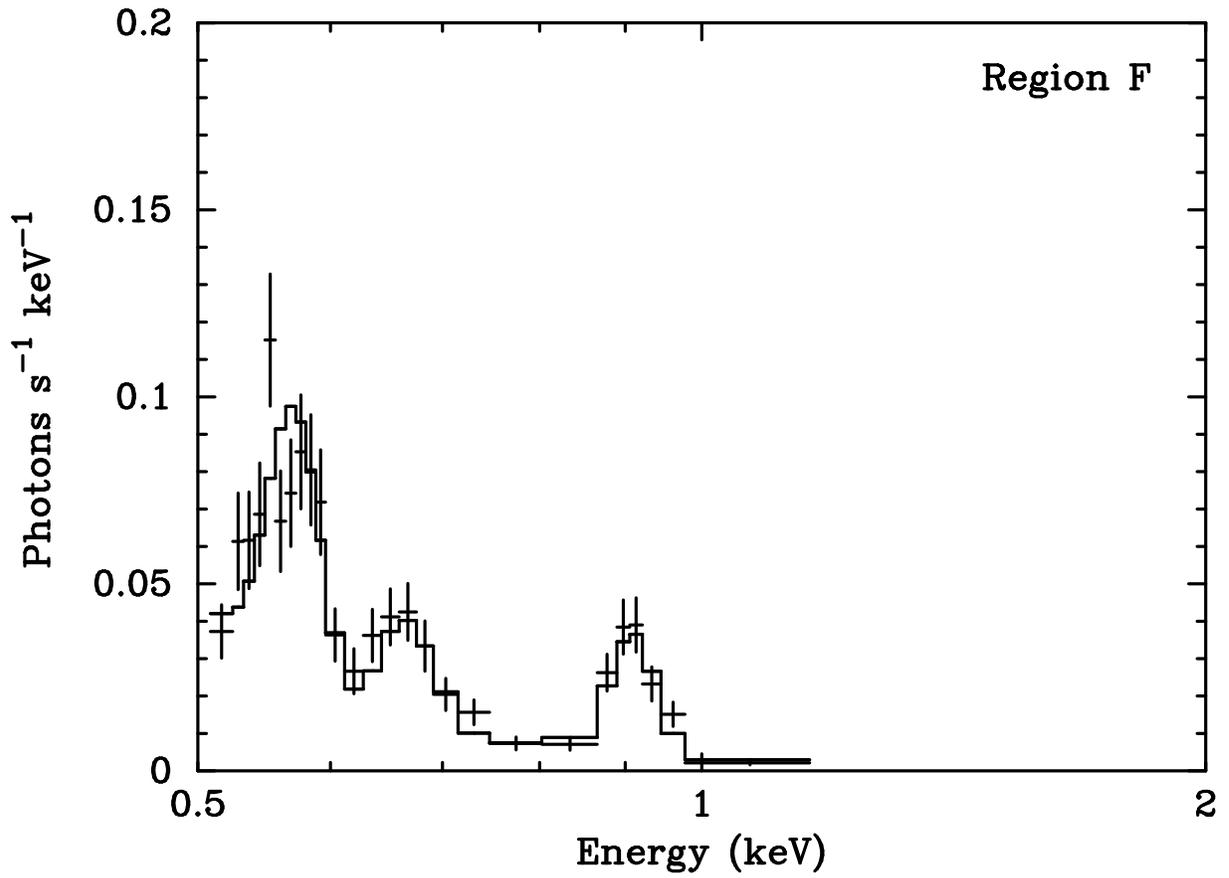} 
\caption{\label{fig:specf}
ACIS spectrum of region F (crosses) with the best-fitting
model (histogram). 
The non-equilibrium ionization plane-parallel shock model
has equilibrium temperature $kT = 0.15$ keV, ionization timescale
$\tau = 2.4\times 10^{12}{\rm \, cm^{-3}\, s}$, and is absorbed
by $N_H = 2.5 \times 10^{21} \psc$.
}
\end{figure}

Globally, we do not expect the interior of the Cygnus Loop to be in ionization
equilibrium, with $\tau \approx 3\times10^{10} {\rm \, cm^{-3}\,s}$
for $n\approx 0.1 {\rm \, cm^{-3}}$ and an age of 8,000 years.
However, the brightest  emission is produced in 
cloud interactions at much higher densities, and are
therefore more likely to be equilibrated.
At the southeast, the fainter (and therefore lower density) edge regions, in which
the elapsed time is also shorter, are most likely to be out of
ionization equilibrium.
In the present observations, non-equilibrium
planar shock models (the VPSHOCK model in XSPEC) improve the spectral
fits of regions A, C, F, and J compared with the corresponding equilibrium
models.  In these cases, we include the integrated contribution from
the time of initial shock ($\tau_i = 0$), and fit for the current value of
the ionization timescale.  
These results are listed in Table \ref{tab:specvp}, 
and Figure \ref{fig:specf} illustrates the non-equilibrium ionization example
of the region F spectrum.

Regions A and C are not strongly
out of equilibrium, with upper limits of  $\tau \approx 5\times10^{12} {\rm \, cm^{-3}\, s}$.
The non-equilibrium effects are stronger in the
lower surface brightness (and therefore likely lower density) regions F and J.
The best-fitting temperature of region J is unphysically high, 
corresponding to an equilibrium 
shock velocity $v_s = 500 {\rm \, km\, s^{-1}}$, 
but it is not constrained well.  The uncertainties encompass
temperatures  that are more typical of the southeast Cygnus Loop
($kT \sim 0.17$ keV).

\begin{figure}
\includegraphics[width=\textwidth]{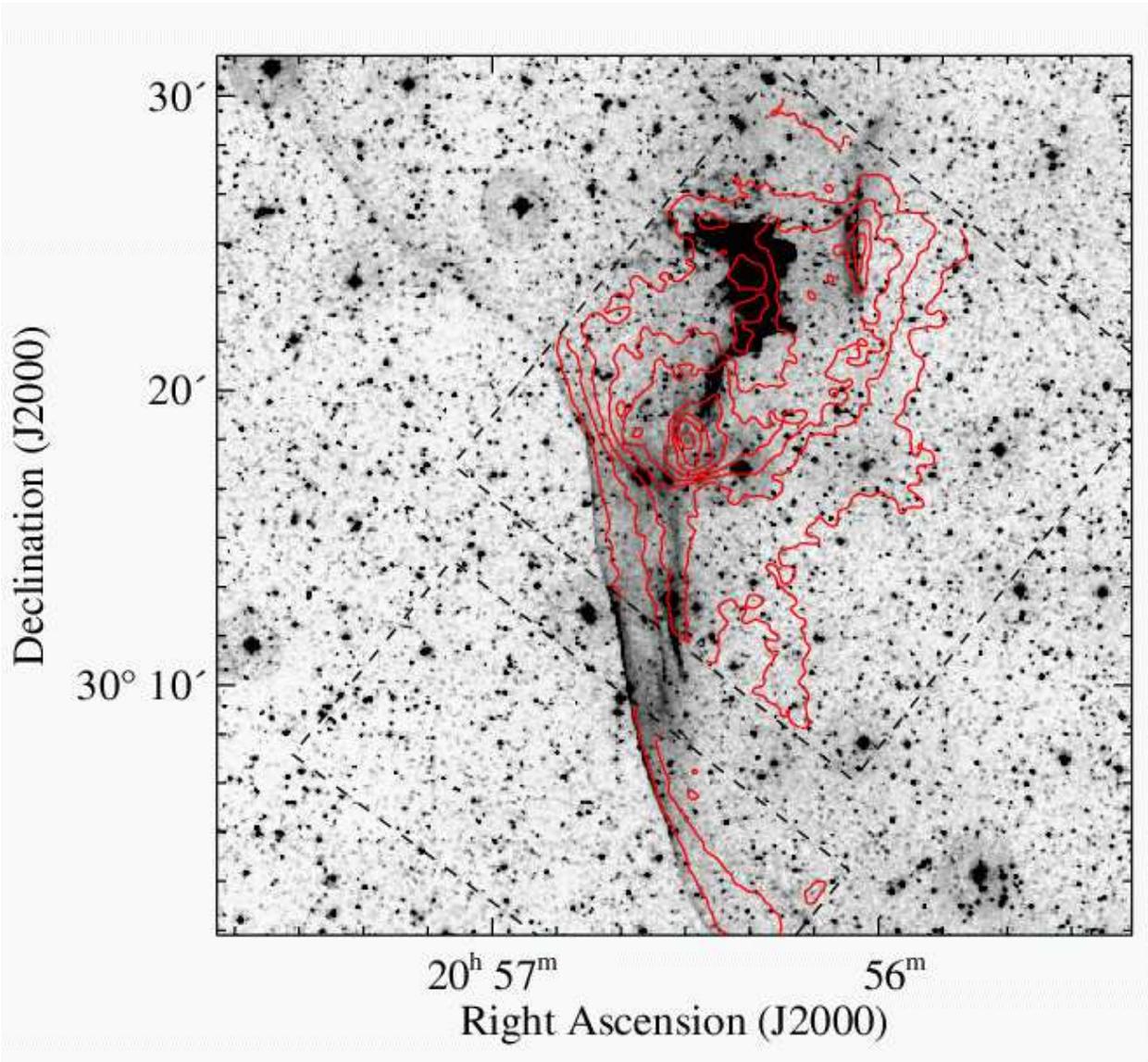} 
\caption{\label{fig:ha}
\chandra{} total-band  contours overlaid on the 
H$\alpha$ image of the southeast knot \citep[from][]{Lev98}.
The optical image is scaled logarithmically, and the 
X-ray contours are linear.  Dashed lines mark the boundary of 
the ACIS field of view.
The brightest X-ray emission (Region A)
is located south of the optically-bright cloud.  
The  X-ray enhancement of Region B is coincident
with a radiative shock.
The narrow H$\alpha$ filaments toward the exterior of the SNR mark the
boundary of the blast wave.
}
\end{figure}

In an effort to measure changes in the emission
behind the shock front and evolution toward equilibrium,
we examined the  spectra 
near the edge of the SNR.
The Balmer-dominated filaments directly reveal the current (projected)
location of the blast wave.
The region around 
$\alpha = 20^{\rm h} 56^{\rm m} 40^{\rm s}, 
\delta=30^\circ 15\arcmin 10\arcsec$ 
offers the clearest signature, with a bright Balmer filament 
marking the shock front (Figure \ref{fig:ha}).
These filaments are not curved, and this region has no other
indication of complicating
projection effects over nearly 2\arcmin{} toward the SNR's interior,
where another nonradiative filament is apparent.

We extracted spectra from four regions 
located at incrementally increasing distance
behind the blast wave (L1--L4).  The regions are relatively narrow
($24\arcsec\equiv 1.6\times10^{17}$ cm) and
extend 170\arcsec{} parallel to the shock front.
We considered fitting the spectra individually and together, to explicitly
follow the evolution toward equilibrium of the shocked material.
In fitting the spectra jointly,
we constrained the column density, abundance, and
equilibrium temperature to be the same in all cases.
The upper bound on ionization timescale is a free parameter
for each region, and this value 
served as the lower bound on $\tau$ for the adjacent
interior region.
We allowed the normalization to vary independently in the four spectra.

The non-equilibrium models do not improve the spectral fits in any case.
The independent fits tend toward equilibrium, 
with $\tau \ge 10^{12} {\rm \, cm^{-3}\, s}$.
Furthermore, in both the independent and the combined fits, 
the ionization parameter does not
vary smoothly behind the blast wave.
These results are surprising because 
the time since shock passage is minimized and the densities
are not necessarily extremely high at these
edge regions, so non-equilibrium conditions would seem likely.
The faintness of the easternmost regions may preclude significant
measurement of their genuine non-equilibrium conditions.
The equilibrium models do show systematic temperature 
variation from the edge to the 
interior, which we discuss in terms of electron-ion temperature
disequilibrium or shock deceleration.

\subsection{Electron-Ion Temperature Disequilibrium\label{subsec:nei}}
Immediately following the passage of a strong shock, 
all particles acquire the same velocity distribution.
Because their masses are different, electrons and protons
are initially heated to different temperatures.  
The temperatures equilibrate through Coulomb collisions in
the post-shock region.
However, plasma instabilities may promote rapid equilibration at the 
shock front \citep{Car88}.

\begin{figure}
\includegraphics[angle=270,width=\textwidth]{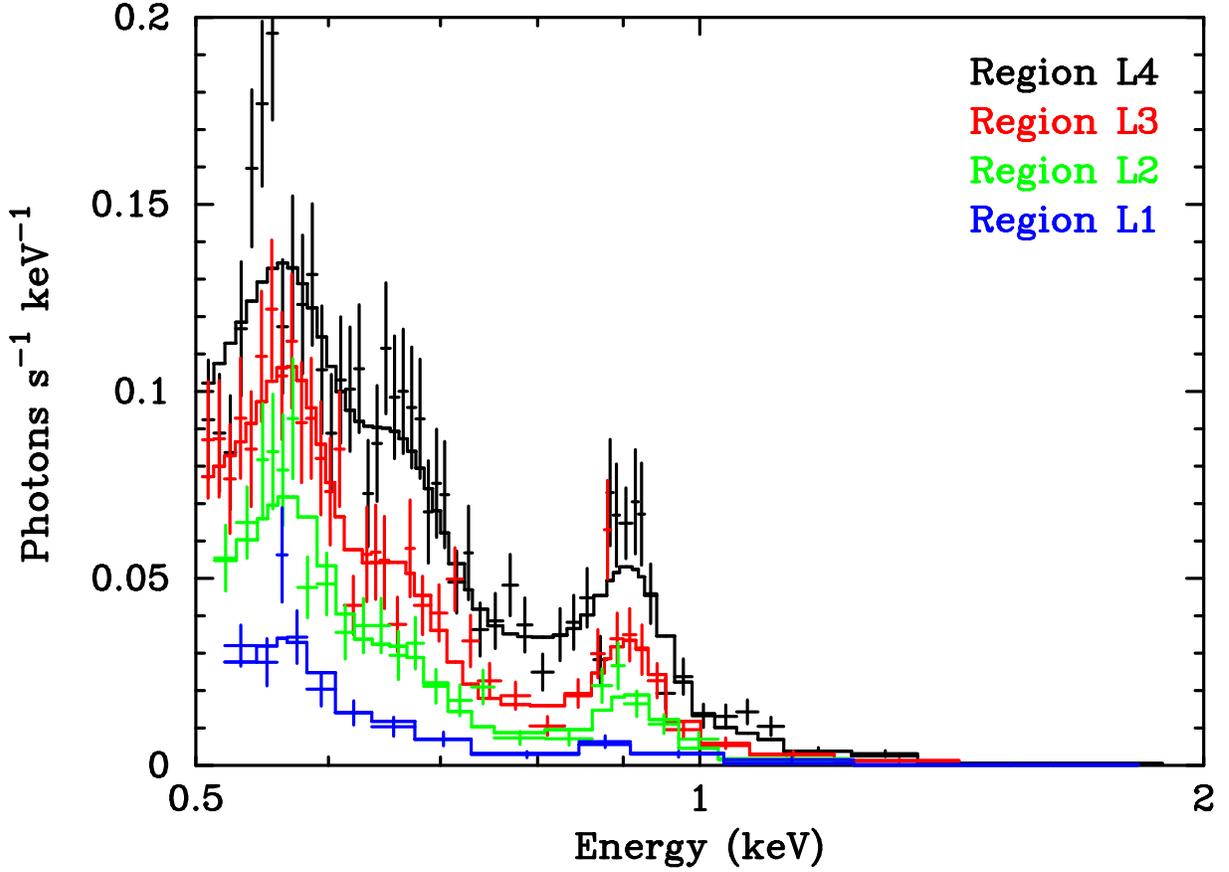} 
\caption{\label{fig:specl}
ACIS spectra (crosses) of regions L1 through L4.
Region L1 (blue) is located $2\times 10^{17}$ cm
behind the blast wave.  Regions L2, L3, and L4
(green, red, and black, respectively)
are each offset a further $2\times 10^{17}$ cm
behind the preceding region.
The column density and abundance are the same in all regions,
while temperature and normalization vary in
the ionization equilibrium models (histograms).
The observed temperature gradient is inconsistent with
Coulomb collisions alone to equilibrate the electron and ion temperatures
behind the shock front.
}
\end{figure}

To investigate the role of collisionless equilibration,
we examine the 
variation of the electron temperature behind the shock front
in the spectra of regions L1--L4.
Although we do not measure the ion temperature, it has
little direct effect on these spectra.
Similar to the combined fits above, 
we simultaneously fit the spectra of regions L1--L4 with the
VEQUIL model 
(Figure \ref{fig:specl}), 
constraining the column density
and abundance to be the same in all cases while allowing the temperature
and normalization to be free parameters.
A single-temperature plasma model provides a reasonable fit, with
sub-solar oxygen abundance and low foreground column
density (Table \ref{tab:specljoint}).
The results are not significantly different from
the independent fits of these regions (Table \ref{tab:spec}),
but applying the common column density and abundance is
more appropriate in this interpretation of the evolution of
a single shock front.

Similar to the independent equilibrium fits of these four spectra,
we find a gradual temperature gradient increasing toward the interior
of the SNR.
In detail, however, 
these measurements disagree with the predictions of
Coulomb equilibration alone.
For a simple comparison, we use  
the observed  Balmer filament to define the location of the blast wave,
and the interior $kT = 0.17$ keV yields the equilibrium electron temperature.
In terms of $T_i$, the ion temperature, the electron
temperature evolves 
$${dT_e\over dt} = {T_i - T_e\over t_{eq}(T_e, T_i)}$$
\citep[cited in \citealt{Spi62}]{Spi40}.
The timescale, $t_{eq}$, is inversely proportional to density,
and for $n = 0.1 {\rm \, cm^{-3}}$, $t_{eq}\approx 10^{11}$ s
as equilibrium $T_e = 1.9\times10^6$ K  is approached.
Assuming the interior region L4 is fully equilibrated, we
use its temperature
to determine the shock velocity 
($v_s = 360 \kms$), and we then relate the elapsed time to the observable
post-shock distances.
For $n = 0.1 {\rm \, cm^{-3}}$ and Coulomb equilibration alone, we would expect
$kT = 0.05$ keV near the edge, and a maximum of only
0.09 keV at the innermost location.  This theoretical
prediction sharply contrasts the observed higher
temperatures and their more gradual rise.

One reasonable concern is that an equilibrium model is used to determine the
electron temperature of each of the regions, while
nonequilibrium ionization would be expected to influence
the effective temperatures we measure.
Over most of the range of timescales and temperatures we observe, however, 
the majority of the line emission is due to \ion{O}{7} transitions.
The absence of higher ionization states does not strongly alter these spectra.
Other indications of non-equilibrium, such as
diagnostic line ratios, cannot be distinguished at the
present low spectral resolution, so they do not drive the spectral fitting.
Furthermore, as described above,  the non-equilibrium ionization models
empirically fail to provide a physically reasonable 
description of the data.
Finally and most significantly, the net effect of 
non-equilibrium ionization would be to {\em increase} the measured temperatures,
requiring yet more rapid  temperature equilibration.

Considering the uncertainties in the temperature measurements,
Coulomb equilibration could be marginally consistent if the density were very high
($n = 2 {\rm \, cm^{-3}}$).  However, the observed flux
precludes such a large density.
The $0.4 \times 3.0 {\rm \, pc^2}$ region L1 is offset $26\arcsec$ 
to the interior of the blast wave.
The nearly-spherical SNR has
a radius of $84 \arcmin$, so  the line of sight through
shocked material at the location of L1 is  $6.8\times 10^{18}$ cm.
From the observed emission of this volume,  we find 
$n = 0.1 {\rm \, cm^{-3}}$. 
The Coulomb equilibration model assumes constant density over
the shocked region, yet region L1 is most likely
the {\em highest} density of the four areas.  It is located
at the boundary of the cavity that surrounded the progenitor
rather than in the evacuated interior.
We therefore conclude that 
Coulomb equilibration is insufficient to produce the
observed variation of electron temperature as a function of 
distance behind the shock front, and rapid post-shock equilibration
is required.  

These results agree with other
observations of similar shocks elsewhere in the Cygnus Loop,
where the initial ion and electron temperatures are 
fully equilibrated
\citep{Gha01,Ray03}.  In general,  
equilibration appears to decrease with increasing shock velocity \citep{Ray03,Rak03}.
The blast wave of the Cygnus Loop is relatively slow, 
in contrast to faster shocks of younger SNRs, which do show
significant temperature disequilibrium between electrons
and ions \citep[e.g.,][]{Ray95,Lam96}. 

The overall decline in temperature toward the SNR's boundary
may be a consequence of global blast wave deceleration in the
surrounding material.
This shell contrasts with
the rarefied bubble that the progenitor evacuated, through
which  the blast wave had propagated previously.
This blast wave deceleration is a genuine dynamical effect,
not merely electron-ion temperature disequilibrium, and it
has been observed  around the entire rim of the Cygnus Loop
\citep{Lev98}.
Adopting the equilibrium temperature  found in region L1
at the edge yields a current blast wave velocity 
$v_{bw} = 330 (+12, -17)\kms$.  This value is somewhat lower
than the shock velocity of the nearby interior region, L4.
A single, abrupt step of density contrast 1.4 could produce this
velocity ratio in the two regions.

\section{Interpretation\label{sec:interp}}

Data from {\it ROSAT} demonstrate that the southeast knot is the result of
interaction of the supernova blast wave with a large-scale
interstellar cloud \citep{Gra95}.  
Here the X-ray rim  is indented from the 
near-circular boundary that can be traced over most
of the SNR's periphery (Figure \ref{fig:hri}).  
Preventing any portion of the projected
blast wave edge from reaching this circular boundary
requires a large obstacle that extends parsecs along the line of sight
to impede the blast wave.  
During the early stages of the encounter,
X-ray emission increases near the interaction site, 
a consequence of pressure enhancement when a reflected
shock propagates back through the previously shocked and compressed
SNR interior.  If the speed of the forward shock propagating through the 
cloud remains high, the
increased density in the cloud shock may also enhance X-ray emission.

The small size of the optical southeast knot 
mistakenly led 
\citet{Fes92} and \citet{Kle94} to envision a physically
small, engulfed cloud.
This early work also  misinterpreted  an optical feature (near region B) as 
the Mach disk that develops after the blast wave has passed.
As \citet{Gra95} show, however, the shock front propagates 
through a significantly neutral, and therefore unshocked medium here.
This analysis---combined with the location of
the X-ray enhancement west of the optical emission,
which requires that the blast wave propagate
east---demonstrates that the optical feature {\em cannot} be  the Mach disk.

These \chandra{} observations further
support the basic interpretation
of the southeast knot as the early stage of interaction between the blast wave
and a large-scale cloud. 
A small, engulfed cloud would not produce the extended
and significant X-ray enhancements that are observed
interior of the optical cloud, and the Mach disk would not
be so luminous. Also, material stripped
from such a cloud would not generate the bright X-rays
toward the exterior at a similar temperature.
We interpret the present results in the context of
interactions with the large cloud and dense material
located at the boundary of the SNR.

While a large obstacle is present,
the X-ray spectral variations indicate that its surface is not
uniform.
The X-ray emission shows significant multiple projections on sub-parsec scales,
similar to the very small scale structure within the optical knot that
high-resolution optical images show \citep{Lev01}.
Regions B and C are sites at the immediate edges the cloud-shock interaction,
with increased X-ray emission located adjacent to the optical emission
of cloud shocks.

Spectrally, 
Region C is hotter  than nearly all the other
regions fit with a single-temperature  model,
independent of whether
the equilibrium or non-equilibrium results are adopted.
Because of the confusion along the line of sight in the diffuse interior
(regions H and I, especially), region L4 offers the best measurement
of the undisturbed blast wave for comparison.  
It is cooler than region C, 
with $kT = 0.17$ keV.
At region C, the greater intensity combined with higher temperature
is consistent with the hotter contribution due to 
a reflected shock.
This characteristic temperature increase would  not be
observed during 
the late stage of interaction (after
the blast wave has completely engulfed a cloud).
A second component of the early interaction, the forward shock
that propagates through the dense cloud, is not directly detected
in the spectrum,
likely because the shock front is rapidly decelerated below
X-ray-emitting temperatures.  
However, this immediate interaction
site does include a softer component that appears as mottled red patches 
in the color image.  These soft X-rays within aperture C contrast with
the exclusively hotter material that extends to the west, 
evident as a bright blue/green area.  Spectrally we find $kT = 0.193\pm0.005$ keV
in the western area, without much emission below 0.6 keV.

In region B,
we would also expect to find evidence of the hotter reflected shock  
or,  if the eastward-moving cloud shock remains fast enough,
the continuation of the slower forward shock.
Surprisingly, the spectrum of region B does not exhibit
a  high temperature similar to region C.
It is somewhat but not significantly hotter than
region L4.
The column density measured at B is higher than that of all  other regions
fit with single-temperature equilibrium models.  
The temperature does increase ($kT = 0.181\pm 0.004$ keV)
if the column density is fixed at a value more typical of the
surrounding areas ($1\times10^{20} \psc$), but the
fit is worse ($\chi^2/\nu = 116/69$).
Alternatively, 
a weaker hot component ($kT = 0.20$ keV) may be included
in addition to the dominant $kT = 0.18$ keV component and
high 
column density,
but it does not significantly improve the fit.
Without definitive spectral identification, 
the X-ray morphology and the association with optical emission
provide the strongest evidence that region B represents
the interaction between the blast wave and a facet of the
large interstellar cloud that is obvious at optical wavelengths.
The narrow area  of high surface brightness distinguishes region B
from the diffuse SNR interior or a projected edge of the undisturbed
blast wave.
This interaction interpretation contrasts alternative
accounts of this  feature as
either the Mach disk of a prior shock or
the emission from a dense
shell that the SNR swept up while propagating through the homogeneous ISM.

Region D
is coincident with
the optical cloud, in which shocks are slowed to
$v_s < 200 \kms$ 
and therefore do not produce X-rays.  
This region shows little or no additional intrinsic absorption, so
the X-ray-emitting region
is likely located in the foreground of the cloud along the same
line of sight.
The volume of hot gas is therefore smaller
than the full line of sight through the SNR, 
resulting in a lower surface brightness.

The highest surface brightness area we observe is region A, 
yet a single-temperature model fits the spectrum well.
(As in region C, the non-equilibrium results are similar
and do not strongly rule out equilibrium conditions.)
Here multiple projections can account for the increased
flux, with a greater column of hot gas located
along a single line of sight.  
This bright spot is located near the apex of two 
nonradiative shocks. 
The Balmer filaments mark
two separate projections of the blast wave, each appearing
when the curved blast wave is viewed at a tangency, through a
long line of sight, so we expect the X-ray surface brightness
to double behind them.  The observed surface brightness is 
increased by a factor of two relative to the immediate surroundings.
Region A is in fact spectrally most similar to region K, which offers
a clear edge-on view immediately behind a single Balmer filament.
The surface brightness of region A is approximately 3 times the
surface brightness of region K.

\begin{figure}
\includegraphics[width=0.85\textwidth]{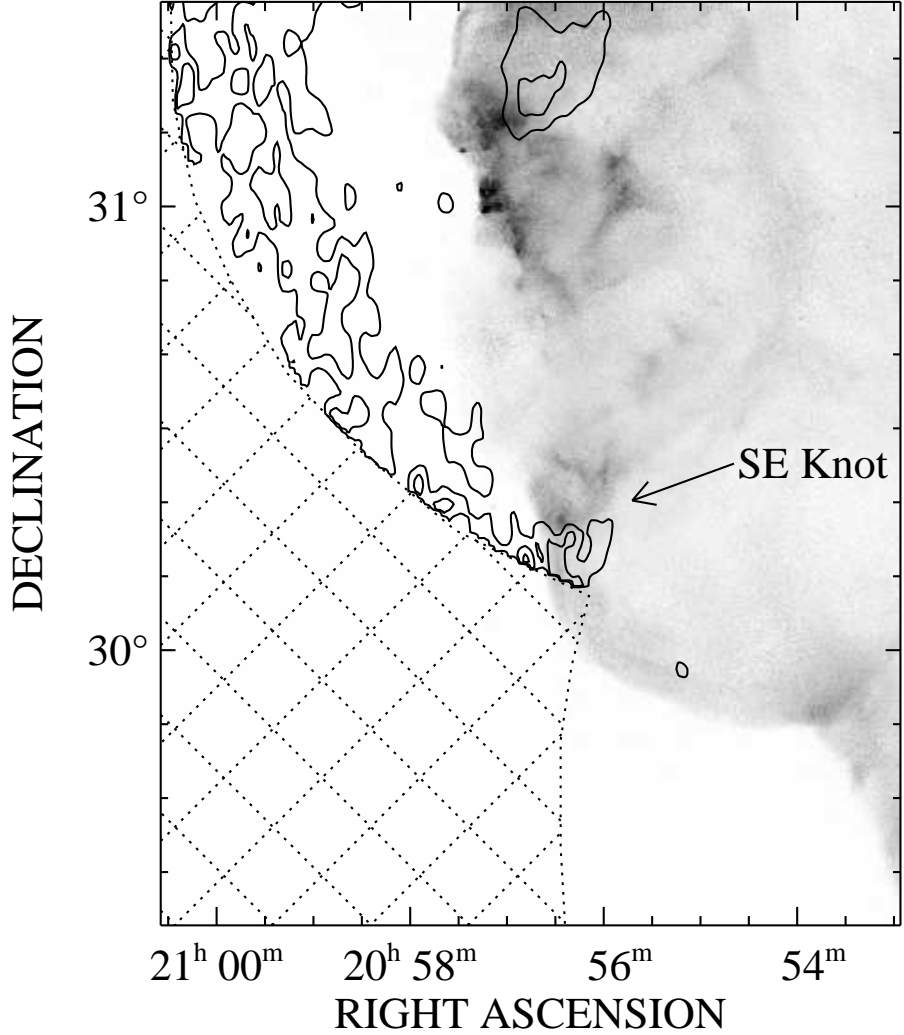} 
\caption{\label{fig:radio}
Contours of neutral hydrogen 21 cm emission \citep[from][]{Lea02}
overlaid on the {\it ROSAT} HRI image.
Neutral material extends up to the southeast X-ray knot but survives immediately
outside, indicating that the blast wave has not fully progressed
(in all projections) far beyond the X-ray knot.
Farther north, \ion{H}{1} emission is offset east of
the current blast wave location, which the X-rays indicate.
Contour levels are $N_{HI} = 2.0$, 2.3, 2.7, and 3.0 $\times 10^{20}\psc$,
in the $2.64$-km-s$^{-1}$-wide band centered on $v_{LSR} = 3.2 \kms$.
Only warm, not cold, neutral material produces the 21 cm emission,
so these contours are lower limits on the total hydrogen column
density.
The angular resolution  of the 21 cm observations is 
$2\arcmin$ (E-W)$\times 4\arcmin$ (N-S), and they
do not cover the area marked with dotted lines.
}
\end{figure}

Observations at 21 cm \citep{Lea02} show the distribution
of warm atomic hydrogen near the southeast knot 
and also indicate that shocks have passed just beyond the X-ray knot.
On large (degree) scales, 
the neutral hydrogen is distributed broadly around the outside of the
SNR.  At the southeast knot, \ion{H}{1} is adjacent to 
the bright X-ray emission, but does not overlap it (Figure \ref{fig:radio}).
This morphology  suggests that a portion of the projected blast wave has
just passed the X-ray knot, and the complete three-dimensional extent of the
blast wave is not located far outside the X-ray knot.
Farther north, the atomic hydrogen is offset to the east of
the SNR blast wave, which the X-rays mark.  This separation
indicates that the neutral surrounding medium 
existed prior to the supernova, and it is not a shell of
ambient material the SNR swept up.
While the distance to the atomic material is uncertain,
the radio emission observed in this velocity range
over the whole Cygnus Loop  is strongly correlated with
optical and X-ray features that are characteristic of interactions
of the blast wave with interstellar clouds,
which indicates that this observed neutral medium and the SNR
are genuinely related \citep{Lea02}.

The connection with the surrounding atomic material suggests 
that the X-ray- and optically-bright areas are the tip of the 
much larger southeast cloud.  The cloud, detectable in \ion{H}{1},
may also have a substantial cold or molecular component.
We estimate the lower limit on the
atomic density very roughly. 
Over $1^\circ$ scales, 
$N_{HI} \approx 2 \times 10^{20} \psc$.  For a line of sight depth
of $1^\circ$, we find $n_{HI} \approx 8 {\rm \, cm^{-3}}$.

\begin{figure}
\includegraphics[width=0.9\textwidth,clip]{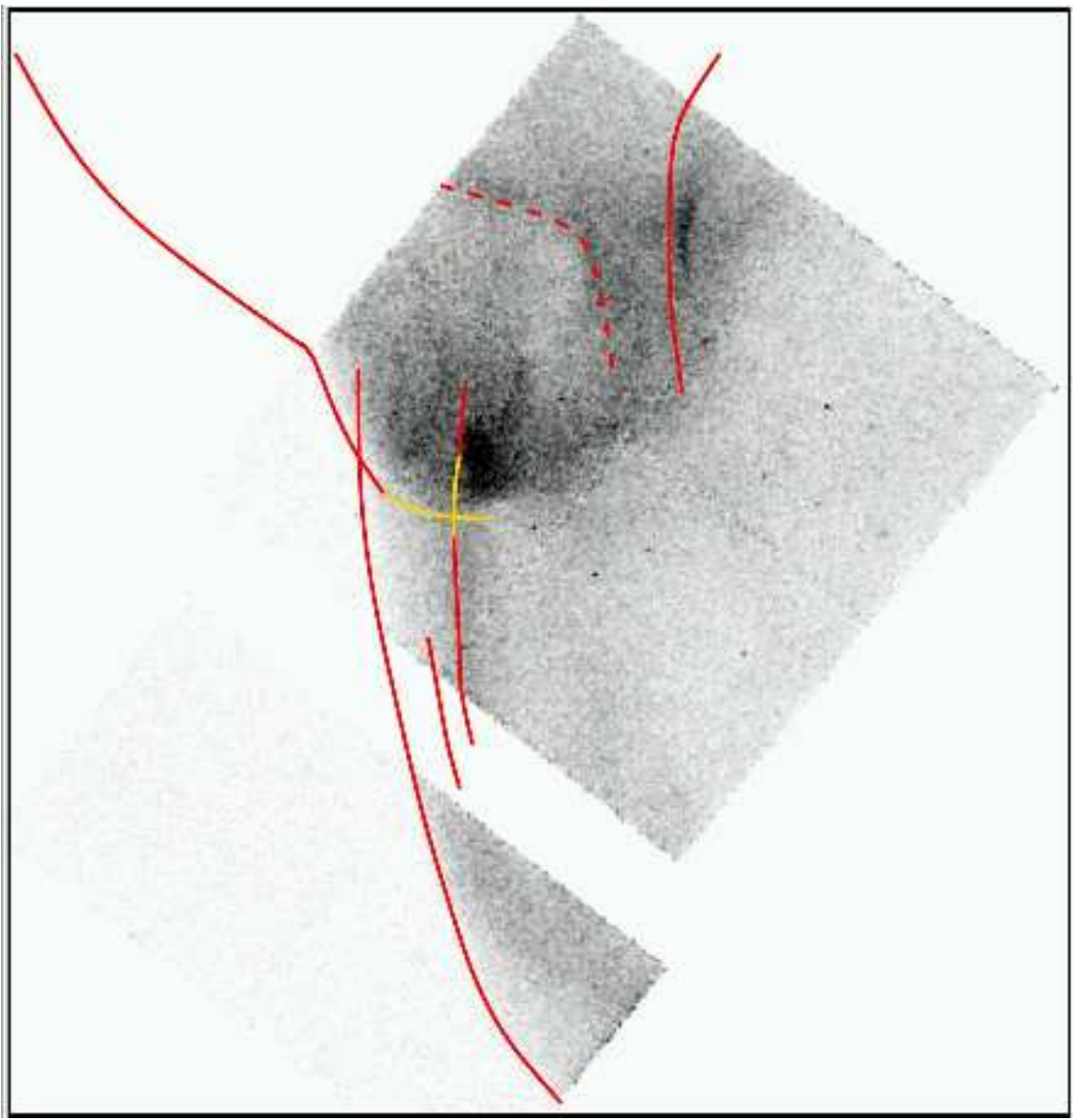} 
\caption{\label{fig:cartoon}
Projected shock front locations on the total-band image, scaled
linearly from 0 (white) to 
$5\times 10^{-7}{\rm \, photons\, cm^{-2}\, arcsec^{-2}}$ (black).
Portions of the projected shock front that are observed as
Balmer filaments are drawn with solid red lines.
The dominant high-temperature component measured in region C
provides evidence for the shock front that
interacts with the optical cloud; 
we identify the extended shock front (dashed red line)
based on the similar total-band and color images.
Overlapping projections of multiple shocks account for
the brightest---yet single-temperature---area, indicating
that the observed shock fronts continue into the projected
interior of the SNR, marked in yellow.
}
\end{figure}

The H$\alpha$ image and \chandra{} spectroscopy together show
the location of many shock front projections.  We illustrate
these schematically in Figure \ref{fig:cartoon}.
The easternmost Balmer filaments are long, arising where the 
blast wave is relatively undisturbed over large distances.
The absolute projected edge is significantly indented relative to
the near-spherical boundary of the Cygnus Loop, so  a
large obstacle that extends several parsecs along the line of sight impedes the
blast wave.
The filaments in the projected interior of the SNR are shorter and
produced near smaller-scale inhomogeneities, which do not preserve
a tangent view through the blast wave over large distances.

The Balmer filament near the cloud interaction of region B is
offset forward (east) of the X-ray enhancement.  The X-rays are
due in part to the shock reflected at region C, which  moves west.
No Balmer filament is associated with the cloud shock of region C.
The X-ray image illustrates that this shock front is extremely
curved.  Maintaining a large line of sight through the edge of the blast
wave while it is strongly deflected around an obstacle is unlikely.

\section{Conclusions}
These \chandra{} observations of the southeast knot of the Cygnus Loop
reveal details of the encounter between
the supernova blast wave and a large interstellar cloud.
Most of the X-ray flux arises in the interaction,
with the increased density of the cloud material and reflected shocks
that further compress and heat previously-shocked gas.
The diffuse interior, through which the 
undisturbed blast wave passed, 
exhibits significant higher-energy emission,
but having lower density, it provides less of the 
total flux within the field of view.

On this small scale, as throughout this SNR as a whole,
the environment determines the X-ray appearance of the Cygnus Loop.
The  complicated   morphology is fundamentally a consequence of geometry,
with significant multiple projections along the line of sight.
Overall, the X-ray spectra are simple and generally indicate
collisional equilibrium in the post-shock regions, even when
the time elapsed since the passage of the blast wave is short.
Only two regions show strong evidence for non-equilibrium ionization,
and electron-ion temperature equilibration behind the shock front
is rapid. 
Optical and radio data further support 
interpretation of the observed spectral variations in terms of 
extrinsic properties of the environment.
Thus, in accounting for the complex appearance of the southeastern
Cygnus Loop, 
we do not find evidence for complex shock physics, such as fluid
instabilities, or significant
intrinsic variations in the evolution of the blast wave.
Higher resolution spectroscopy at similar spatial
resolution could yet reveal such complex physics,
but these data clearly show significant spectral variations
on small spatial scales, which we attribute to the SNR's environment.

\acknowledgements
We thank the referee, P. Slane, for suggestions that significantly
improved this work.
We thank D. Leahy for providing an electronic version of the \ion{H}{1} map.
This research was supported by Chandra grant GO0-1121 and NSF CAREER 
award AST-0237291.

\begin{deluxetable}{lllllrc}
\tablewidth{0pt}
\tablecaption{Single-Temperature Spectral Models\label{tab:spec}}
\tablehead{
\colhead{Region}
&\colhead{$N_H$\tablenotemark{a}}
&\colhead{$kT$\tablenotemark{b}}
&\colhead{$A$\tablenotemark{c}}
&\colhead{$Z_O/Z_\sun$}
&\colhead{Counts}
&\colhead{$\chi^2$/dof}
}
\startdata
 A & $ 1.0\pm0.3$ & $0.168\pm0.003$ & $  6.1^{+1.2}_{-1.0}$ & $0.35^{+0.05}_{-0.04}$ & $5621$ & $ 78/74$\\
 B & $ 1.6\pm0.3$ & $0.175\pm0.005$ & $  2.9^{+0.8}_{-0.6}$ & $0.41^{+0.08}_{-0.07}$ & $2660$ & $106/68$\\
 C & $ 0.9\pm0.3$ & $0.191\pm0.005$ & $  1.8\pm0.4$ & $0.51^{+0.08}_{-0.07}$ & $3736$ & $ 98/72$\\
 D & $ 0.8^{+0.3}_{-0.3:}$ & $0.188\pm0.005$ & $  1.7\pm0.4$ & $0.36^{+0.06}_{-0.05}$ & $3059$ & $102/69$\\
 E & $ 1.2^{+0.5}_{-0.4}$ & $0.169^{+0.005}_{-0.006}$ & $  2.0^{+0.8}_{-0.5}$ & $0.45^{+0.13}_{-0.09}$ & $2029$ & $ 56/55$\\
 F & $ 0.8^{+1.0}_{-0.3:}$ & $0.149^{+0.008}_{-0.015}$ & $  1.1^{+1.4}_{-0.3}$ & $0.37^{+0.18}_{-0.09}$ & $ 693$ & $ 37/22$\\
 G & $ 2.5\pm0.2$ & $0.182^{+0.003}_{-0.004}$ & $  6.8^{+1.2}_{-1.0}$ & $0.51^{+0.06}_{-0.05}$ & $4981$ & $173/86$\\
 H & $ 2.3^{+0.3}_{-0.2}$ & $0.189\pm0.004$ & $  5.7^{+1.1}_{-0.9}$ & $0.57^{+0.07}_{-0.06}$ & $5410$ & $188/92$\\
 I & $ 1.8\pm0.3$ & $0.178^{+0.003}_{-0.004}$ & $  5.2^{+1.1}_{-0.8}$ & $0.43^{+0.07}_{-0.06}$ & $4935$ & $188/88$\\
 J & $ 0.5^{+0.6}_{-0.0:}$ & $0.164^{+0.004}_{-0.007}$ & $  1.8^{+0.9}_{-0.1}$ & $0.32^{+0.08}_{-0.05}$ & $2128$ & $ 69/51$\\
 K & $ 1.2\pm0.4$ & $0.165^{+0.004}_{-0.005}$ & $  3.8^{+1.1}_{-0.8}$ & $0.38^{+0.09}_{-0.07}$ & $3095$ & $ 62/69$\\
L1 & $ 1.4^{+2.6}_{-0.9:}$ & $0.12\pm0.03$ & $  1.4^{+20}_{-1.4:}$ & $0.38^{+0.98}_{-0.23}$ & $ 294$ & $ 10/9$\\
L2 & $ 0.5^{+4.0}_{-0.0:}$ & $0.145^{+0.007}_{-0.06}$ & $  1.0^{+140}_{-0.1}$ & $0.31^{+0.09}_{-0.08}$ & $ 653$ & $ 27/20$\\
L3 & $ 0.5^{+0.8}_{-0.0:}$ & $0.151^{+0.005}_{-0.01}$ & $  1.4^{+1.2}_{-0.1}$ & $0.29^{+0.07}_{-0.05}$ & $1080$ & $ 38/34$\\
L4 & $ 0.8^{+0.6}_{-0.3:}$ & $0.165^{+0.006}_{-0.008}$ & $  1.6^{+0.9}_{-0.4}$ & $0.46^{+0.15}_{-0.11}$ & $1724$ & $ 72/48$\\
\enddata 
\tablenotetext{a}{Column density, in units of $10^{21} {\rm \, cm ^{-2}}$.}
\tablenotetext{b}{Temperature of thermal plasma, in keV.}
\tablenotetext{c}{Normalization of thermal component in units of $10^{-3} K$, 
where $K=(10^{-14}/(4\pi D^2))\int n_e n_H dV,\ D$ is the distance to the source (cm),
$n_e$ is the electron density (${\rm cm^{-3}}$), and $n_H$ is the hydrogen density 
(${\rm cm^{-3}}$).}
\tablecomments{Parameters and errors constrained by hard limits are
marked with a colon.}
\end{deluxetable}

\begin{deluxetable}{lllllllrc}
\tablewidth{0pt}
\tablecaption{Two-Temperature Spectral Models\label{tab:spec2}}
\tablehead{
\colhead{Region}
&\colhead{$N_H$\tablenotemark{a}}
&\colhead{$kT_1$\tablenotemark{b}}
&\colhead{$A_{1}$\tablenotemark{c}}
&\colhead{$Z_O/Z_\sun$}
&\colhead{$kT_2$\tablenotemark{b}}
&\colhead{$A_{2}$\tablenotemark{d}}
&\colhead{Counts}
&\colhead{$\chi^2$/dof}
}
\startdata
 G & $ 1.8^{+0.3}_{-0.5}$ & $0.180^{+0.006}_{-0.004}$ & $  4.4^{+1.1}_{-1.4}$ & $0.48^{+0.06}_{-0.05}$ & $0.55^{+0.05}_{-0.07}$ & $ 8.1^{+2.0}_{-1.8}$ & $4981$ & $130/84$\\
 H & $ 0.5^{+0.6}_{-0.0:}$ & $0.202^{+0.006}_{-0.013}$ & $  1.6^{+0.6}_{-0.2}$ & $0.53^{+0.09}_{-0.07}$ & $0.69^{+0.08}_{-0.06}$ & $ 6.9^{+1.5}_{-1.7}$ & $5410$ & $152/90$\\
 I & $ 0.8^{+0.4}_{-0.3:}$ & $0.175^{+0.006}_{-0.005}$ & $  2.8^{+1.2}_{-0.5}$ & $0.40^{+0.07}_{-0.06}$ & $0.54^{+0.06}_{-0.07}$ & $ 7.5^{+1.9}_{-1.4}$ & $4935$ & $120/86$\\
\enddata 
\tablenotetext{a}{Column density, in units of $10^{21} {\rm \, cm ^{-2}}$.}
\tablenotetext{b}{Temperature of thermal plasma, in keV.}
\tablenotetext{c}{Normalization of thermal component in units of $10^{-3} K$, 
where $K=(10^{-14}/(4\pi D^2))\int n_e n_H dV,\ D$ is the distance to the source (cm),
$n_e$ is the electron density (${\rm cm^{-3}}$), and $n_H$ is the hydrogen density 
(${\rm cm^{-3}}$).}
\tablenotetext{d}{Normalization of thermal component in units of $10^{-5} K$.}
\end{deluxetable}

\begin{deluxetable}{llllllrr}
\tablewidth{0pt}
\tablecaption{Plane-Parallel Shock Models\label{tab:specvp}}
\tablehead{
\colhead{Region}
&\colhead{$N_H$\tablenotemark{a}}
&\colhead{$kT$\tablenotemark{b}}
&\colhead{$A$\tablenotemark{c}}
&\colhead{$Z_O/Z_\sun$}
&\colhead{$\tau$\tablenotemark{d}}
&\colhead{Counts}
&\colhead{$\chi^2$/dof}
}
\startdata
 A & $ 1.9^{+0.5}_{-0.9}$ & $0.174^{+0.007}_{-0.009}$ & $  7.2^{+1.7}_{-1.3}$ & $0.48\pm0.12$ & $ 0.84^{+5}_{ -0.4}$ & $5621$ & $ 74/73$\\
 C & $ 1.7^{+0.7}_{-0.8}$ & $0.198^{+0.02}_{-0.009}$ & $  2.1^{+0.7}_{-0.5}$ & $0.67\pm0.2$ & $ 1.0^{+5}_{ -0.6}$ & $3736$ & $ 94/71$\\
 F & $ 2.5^{+3.1}_{-1.0}$ & $0.15^{+0.03}_{-0.05}$ & $  2.5^{+180}_{-1.5}$ & $0.56\pm0.2$ & $ 0.24^{+ 0.6}_{ -0.2}$ & $ 693$ & $ 28/21$\\
 J & $ 0.5^{+1.5}_{-0.0:}$ & $0.32^{+0.06}_{-0.2}$ & $  0.3^{+2.2}_{-0.1}$ & $0.46^{+0.1}_{-0.07}$ & $ 0.021^{+ 0.34}_{ -0.005}$ & $2128$ & $ 63/50$\\
\enddata 
\tablenotetext{a}{Column density, in units of $10^{21} {\rm \, cm ^{-2}}$.}
\tablenotetext{b}{Temperature of thermal plasma, in keV.}
\tablenotetext{c}{Normalization of thermal component in units of $10^{-3} K$, 
where $K=(10^{-14}/(4\pi D^2))\int n_e n_H dV,\ D$ is the distance to the source (cm),
$n_e$ is the electron density (${\rm cm^{-3}}$), and $n_H$ is the hydrogen density 
(${\rm cm^{-3}}$).}
\tablenotetext{d}{Ionization timescale, in units of $10^{12} {\rm \, cm^{-3}\,s}$.}
\tablecomments{Parameters and errors constrained by hard limits are
marked with a colon.}
\end{deluxetable}

\begin{deluxetable}{rllllrc}
\tablewidth{0pt}
\tablecaption{Combined Edge Region Spectral Model\label{tab:specljoint}}
\tablehead{
\colhead{Region}
&\colhead{$N_H$\tablenotemark{a}}
&\colhead{$kT$\tablenotemark{b}}
&\colhead{$A$\tablenotemark{c}}
&\colhead{$Z_O/Z_\sun$}
&\colhead{Counts}
&\colhead{$\chi^2$/dof}
}
\startdata
& $ 0.5^{+0.4}_{-0.0:}$ &&& $0.35^{+0.07}_{-0.04}$ && $153/117$\\
L1 & & $0.135^{+0.011}_{-0.013}$ & $  0.5^{+0.4}_{-0.2}$ & & $294$ & \\
L2 & & $0.146^{+0.006}_{-0.009}$ & $  0.9^{+0.5}_{-0.2}$ & & $ 653$ & \\
L3 & & $0.153^{+0.005}_{-0.008}$ & $  1.3^{+0.6}_{-0.2}$ & & $1080$ & \\
L4 & & $0.165^{+0.003}_{-0.004}$ & $  1.6^{+0.6}_{-0.2}$ & & $1724$ & \\
\enddata 
\tablenotetext{a}{Column density, in units of $10^{21} {\rm \, cm ^{-2}}$.}
\tablenotetext{b}{Temperature of thermal plasma, in keV.}
\tablenotetext{c}{Normalization of thermal component in units of $10^{-3} K$, 
where $K=(10^{-14}/(4\pi D^2))\int n_e n_H dV,\ D$ is the distance to the source (cm),
$n_e$ is the electron density (${\rm cm^{-3}}$), and $n_H$ is the hydrogen density 
(${\rm cm^{-3}}$).}
\tablecomments{Parameters and errors constrained by hard limits are
marked with a colon.}
\end{deluxetable}

\clearpage

\end{document}